\documentclass[
  reprint,  
  amsmath,amssymb,aps,showpacs,longbibliography,superscriptaddress
nofootinbib, showkeys
]{revtex4-1}
\usepackage{amssymb}
\usepackage{graphicx}
\usepackage{booktabs}
\usepackage{dcolumn}
\usepackage{bm}
\usepackage{mathrsfs}
\usepackage{mathtools}
\usepackage{amsmath}
\usepackage{amssymb}
\usepackage{epstopdf}
\usepackage{comment}
\usepackage{units}
\usepackage{mathtools}
\usepackage{dsfont}
\usepackage{upgreek}
\usepackage[font=footnotesize]{caption}
\usepackage{subfig}
\begin{document}
\preprint{APS/123-QED}

\title{Irreversible dynamics of a continuum driven by active matter}

\author{John C. Neu}
\affiliation{University of California at Berkeley, Department of Mathematics}

\author{Stephen W. Teitsworth}
\affiliation{Duke University, Department of Physics, Box 90305
Durham, NC 27708-0305}

\date{\today}

\begin{abstract}

We study the fluctuational behavior of overdamped elastic filaments (e.g., strings or rods) driven by active matter which induces irreversibility.  The statistics of discrete normal modes are translated into the continuum of the position representation which allows discernment of the spatial structure of dissipation and fluctuational work done by the active forces.  The mapping of force statistics onto filament statistics leads to a generalized fluctuation-dissipation relation which predicts the components of the stochastic area tensor and its spatial proxy, the irreversibility field.  We illustrate the general theory with explicit results for a tensioned string between two fixed endpoints. Plots of the stochastic area tensor components in the discrete plane of mode pairs reveal how the active forces induce spatial correlations of displacement along the filament.  The irreversibility field provides additional quantitative insight into the relative spatial distributions of fluctuational work and dissipative response.

\end{abstract} 

\maketitle

\section{Introduction}
\label{sec:intro}

Noise-driven phenomena associated with non-equilibrium dynamics occur throughout the natural sciences \cite{Gnesotto_2018, Weiss_PRE_2007, Weiss_JSP_2019, Gieseler_2014, Millen_2014,Gonzalez_PRE_2019, Bomze_PRL_2012, Bowick_PRX_2022}.  Such systems include, for example, noise-driven electronic circuits \cite{Chiang_2017a, Chiang_2017b, Gonzalez_PRE_2019, Teitsworth_2019}, climate models \cite{Penland_JClimate_1995, Weiss_JSP_2019, Berner_JClimate_2020}, as well as biophysical systems. An example of the latter is provided by artificial or natural filamentary structures embedded in cytoskeletal networks subject to the fluctuational effects of molecular motors \cite{Gnesotto_2018,  Brangwynne_PRL_2008, Fakhri_Science_2014, Bacanu_2023}.  Traditional metrics for quantifying the departure from equilibrium such as entropy production and heat transfer driven by thermal gradients have the advantages of their direct physical inspirations \cite{Ciliberto_PRL_2013, Ciliberto_JSM_2013b, Gnesotto_2018, Gingrich_2019}.   However, in many systems, especially living matter, thermal gradients are \textit{not} the driving force \cite{Battle_Science_2016, Teitsworth_2019, Bacanu_2023}.   For example, biophysical systems such as beating flagella and cilia are typically driven by chemical (metabolic and enzymatic) processes \cite{Battle_Science_2016}.  Another set of examples is provided by linear and nonlinear electronic circuits which may be driven by external voltage noise sources as well as internal non-thermal sources, for example, shot noise associated with nonlinear elements such as tunnel diodes or quantum wells \cite{Tretiakov_2003, Bomze_PRL_2012, Teitsworth_2019}.   Alternative metrics of irreversibility have recently been developed and successfully applied with a focus on low dimensional systems or low dimensional projections of high dimensional systems.  Such methods include stochastic area \cite{Ghanta2017, Teitsworth_PRE_2022, Touchette_PRE_2023}, cycling frequencies \cite{Mura_2018,  Gradziuk_2019}, as well direct projections of measured probability currents \cite{Battle_Science_2016, Chiang_2017a, Gnesotto_2018}.

In this paper, we develop a theoretical framework for implementing the extension of stochastic area to high dimensional systems such as those associated with continuous degrees of freedom, e.g., a noise-driven string. One motivation for doing so comes from recent experiments on biophysical systems measuring the conformational dynamics of cilia or other mechanical probes embedded in viscoelastic networks and driven by both thermal noise and molecular motors which provide active sources of noise \cite{Bacanu_2023, Gladrow_PRL_2016, Battle_Science_2016, Brangwynne_PRL_2008}.  In such experiments, one typically analyzes time series for just a few state variables, thus sampling low dimensional projections of high dimensional phase space. For instance, Battle \textit{et al.} \cite{Battle_Science_2016} record time series of beating cilia using video microscopy and infer time series of a few normal mode amplitudes.  Statistics in the phase plane of just two normal modes can display the signatures of irreversibility.  In that paper, they construct empirical maps of the probability current in such phase planes, the circulation of which exhibits well-defined elliptical streamlines under non-equilibrium conditions.  In more recent experiments, Bacanu \textit{\textit{et al.}} have studied the fluctuational properties of carbon nanotubes inserted into the living cells \cite{Bacanu_2023}.  Instead of mapping probability currents, they estimate an antisymmetrized two time autocorrelation for various pairs of normal modes.  By the fundamental work of Onsager \cite{Onsager_1931}, the non-vanishing of this antisymmetrized autocorrelation tensor indicates irreversibility.  
 
 In the current paper, we focus on the overdamped fluctuation statistics of an elastic continuum driven by active noise.  The state variable is taken as the continuum limit of one component of the displacement from the equilibrium position.  \textit{Active noise} refers to the statistics of the fluctuational forces acting on the elastic continuum which induces irreversible statistics in the dissipative motion of the latter.   For clarity of a first impression, we assume here that the force statistics by itself is reversible, with a correlation time that is much shorter than the relaxation time of any normal mode of the filament.  
 
 In Section II, we introduce an irreversibility field that serves as a metric of irreversibility and applies to spatially extended elastic continuums.  It is closely related to the \textit{stochastic area} introduced for dynamics with discrete degrees of freedom  \cite{Ghanta2017, Teitsworth_PRE_2022, Touchette_PRE_2023}.  Section III describes the overdamped stochastic dynamics of the continuum object.  Section IV treats the mapping from force statistics to normal mode statistics which can be viewed as a type of generalized fluctuation-dissipation relation.  Following the tradition of current literature, we represent the forces acting on the continuum and induced displacements from mechanical equilibrium by linear combinations of normal modes; the coefficients in these expansions serve as amplitudes. Section V focuses on illustrating general results with explicit calculations for the case of a tensioned string between fixed endpoints serving as the continuum filament.  In analogy with the experimental results of Bacanu \textit{et al.} \cite{Bacanu_2023}, we compute explicit maps of the components of stochastic area in the discrete plane of mode pairs.  These maps show clear dependence on  properties of the embedding medium and the localized nature of the active forces. Section VI translates the amplitude statistics into \textit{spatial} statistics with a focus on the correlations between displacements associated with two distinct positions.  In this way, we are able to discern the spatial structure of where the dissipative motional responses occur relative to the active fluctuations that drive the filament.



\section{The irreversibility and stochastic area fields}
\label{sec:irreversibility}

\begin{figure}
\centerline{\includegraphics[width=0.40\textwidth]{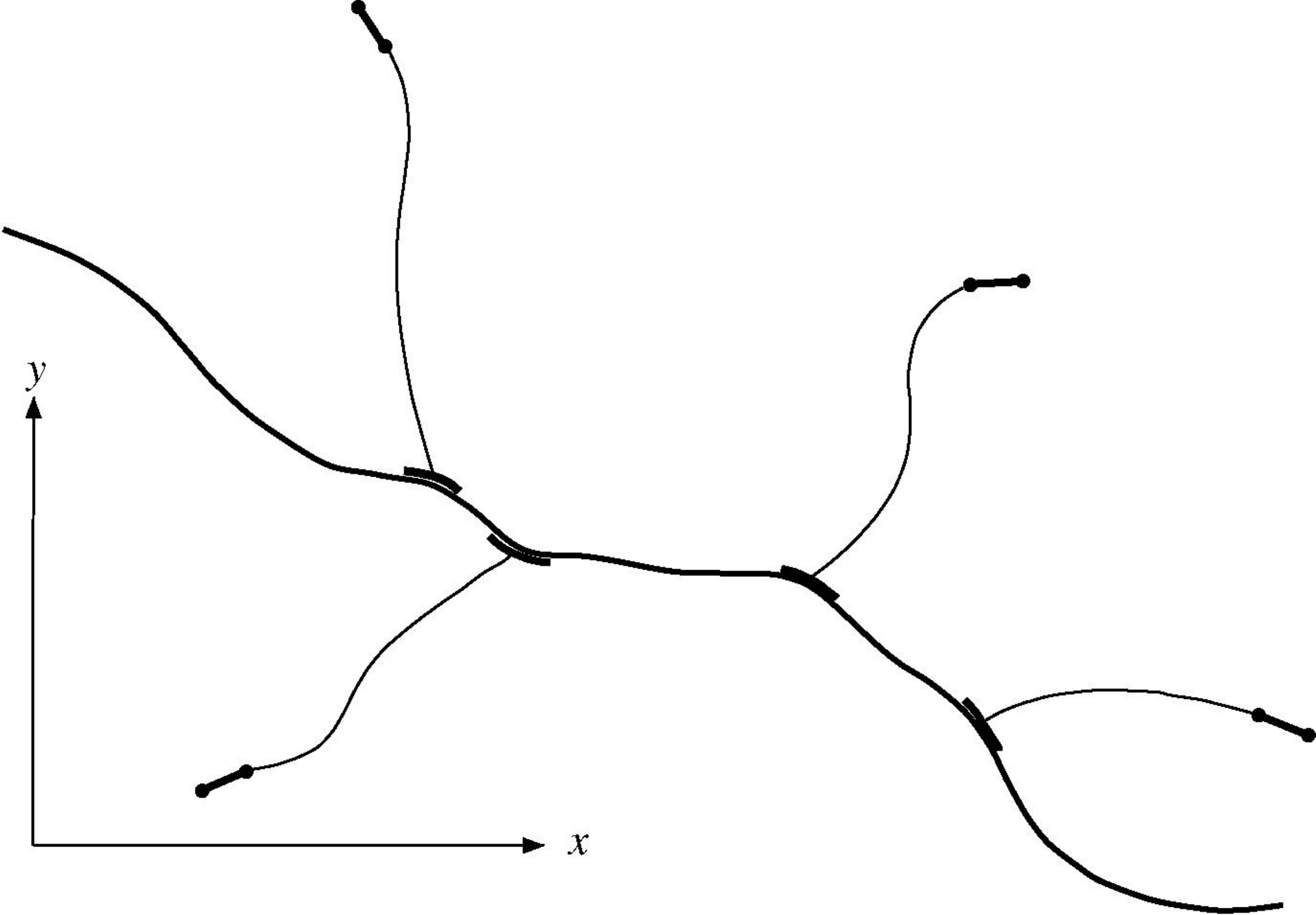}}
\caption{Schematic illustration of a one-dimensional filament embedded in a viscoelasitc network (not shown) and subject to active stochastic forcing from nearby molecular motors.  Motors are represented by dumbbell shaped objects and only parts of the network that connect motors to the filament are shown. The viscoelastic network is assumed to impart equilibrium thermal fluctuations uniformly along the length filament.}
\label{FIG1}
\end{figure}

The rest state of a mechanical continuum is some bounded region $R$ embedded in one-, two- or three-dimensional space.  The scalar field $X(\mathbf{x}, t)$  represents one component of the displacement from the rest position $\mathbf{x} \in R$ as a function of time $t$.  For example, $R$ might denote a one-dimensional contour that corresponds to a filament embedded in a three-dimensional viscoelastic network and driven by active noise as depicted in Figure \ref{FIG1}.  In this case,  $X(\mathbf{x}, t)$ could represent one component of the transverse displacement of the filament from its equilibrium conformation.   

We have a statistical ensemble of such scalar fields and the focus is on \textit{stationary} statistics.  Therefore, the ensemble is invariant under time translation:   For any $\tau$, the ensemble of time translated fields $X(\mathbf{x}, t + \tau)$ is the same as the original.  Following Onsager \cite{Onsager_1931}, the field statistics is 
\textit{microscopically reversible} if the ensemble of fields $X(\mathbf{x}, t)$ is the same as the ensemble of their time reversals $X(\mathbf{x}, -t)$.  Then, denoting ensemble averaging by $\langle \cdot \rangle$, we can write
\begin{equation}
\langle X(\mathbf{x}, t) X(\mathbf{y}, t') \rangle = \langle X(\mathbf{x}, -t) X(\mathbf{y},- t') \rangle.
\end{equation}
  Due to stationarity, the right hand side is unchanged by adding any given number to both time arguments $-t$ and $-t'$.  Taking this number to be $t +t'$, we arrive at
\begin{equation}
\langle X(\mathbf{x}, t) X(\mathbf{y}, t') \rangle = \langle X(\mathbf{x}, t') X(\mathbf{y}, t) \rangle
\end{equation}
This means that, if the correlation function defined by
\begin{eqnarray}
& & \omega (\mathbf{x}, \mathbf{y}, \tau) :=  \nonumber \\ & & \:\:\:\:\:\: \frac{1}{2} \langle X(\mathbf{x}, t+\tau) X(\mathbf{y}, t) - X(\mathbf{x}, t) X(\mathbf{y}, t+\tau) \rangle
\end{eqnarray}
is not identically zero, the filament statistics must be irreversible.  We call $\omega(\mathbf{x}, \mathbf{y}, \tau)$ the \textit{irreversibility field with time lag $\tau$}.
In the limit $\tau \rightarrow 0$,
\begin{eqnarray}
& & \frac{\omega (\mathbf{x}, \mathbf{y}, \tau)}{\tau} \rightarrow \Omega(\mathbf{x}, \mathbf{y}) :=  \nonumber \\ & & \:\:\:\:\:\:  \frac{1}{2} \langle \partial_t X(\mathbf{x}, t) X(\mathbf{y}, t) - X(\mathbf{x}, t) \partial_t X(\mathbf{y}, t) \rangle.
\end{eqnarray}
We call $\Omega(\mathbf{x}, \mathbf{y})$ the \textit{irreversibility field}.  Both $\Omega(\mathbf{x}, \mathbf{y})$ and $\omega(\mathbf{x}, \mathbf{y}, \tau)$ are defined on the space $R \times R$.

The experimental detection of irreversibility may proceed by recording large ensembles of time series of $X(\mathbf{x}, t)$ and $X(\mathbf{y}, t)$ at two points $\mathbf{x}, \mathbf{y}$.  For a given time lag, we can estimate $\omega (\mathbf{x}, \mathbf{y}, \tau)$ by empirical averaging.  Some experimentalists favor a range of finite lag times $\tau$ \cite{Bacanu_2023}.  Here, we show that the limit $\tau \rightarrow 0$ is experimentally feasible.  That is, one can make robust estimates of the irreversibility field $\Omega(\mathbf{x}, \mathbf{y})$ from the same data.

From time series of $X(\mathbf{x}, t)$ and $X(\mathbf{y}, t)$ of the field at two points $\mathbf{x}$ and $\mathbf{y}$, form the parametric curve $C(t)$ in the $u, v$ plane,
\begin{equation}
C(t): \: (u, v) = (X(\mathbf{x}, t'), X(\mathbf{y}, t')), \: 0 < t' <t.
\end{equation}
 The stochastic line integral
 \begin{equation}
 a(\mathbf{x}, \mathbf{y}, t) := \frac{1}{2} \int_{C(t)} u \; dv - v \; du
 \end{equation}
 is called the \textit{stochastic area} swept out by an individual trajectory in the $u, v$ plane in the time interval $(0, t)$ \cite{Ghanta2017, Gonzalez_PRE_2019, Teitsworth_PRE_2022, Touchette_PRE_2023}.  The parametric form of the line integral is

 \begin{equation}
 \label{eq:stoch_area_param_form}
 \begin{split}
a(\mathbf{x}, \mathbf{y}, t) =  
\frac{1}{2} \int_0^t \Big[(X(\mathbf{x}, t') \partial_t X(\mathbf{y}, t') \\
- X(\mathbf{y}, t') \partial_t X(\mathbf{x}, t)\Big] \; dt'. 
\end{split}
\end{equation}

 Ensemble averaging gives
 \begin{equation}
 \langle a(\mathbf{x}, \mathbf{y}, t) \rangle =  -t \Omega(\mathbf{x}, \mathbf{y}).
 \end{equation} 
 
 The nominal experiment is to record a large ensemble of time series $X(\mathbf{x}, t)$ and $X(\mathbf{y}, t)$, and for each compute the stochastic area.  A  nonzero constant growth rate of the empirically averaged stochastic area indicates irreversibility.   Alternatively, let $a(\mathbf{x}, \mathbf{y}, t)$ be a \textit{single} realization of stochastic area in $0 < t < \infty$.  If the correlation time of the field is finite, we anticipate that the time average reproduces the ensemble average \cite{Ghanta2017, Touchette_PRE_2023}:
 \begin{equation}
\lim_{T \rightarrow \infty}  \frac{1}{T} \int_0^T a(\mathbf{x}, \mathbf{y}, t) \; dt = -\Omega(\mathbf{x}, \mathbf{y}).
 \end{equation}
We note that, in general, the stochastic area $a(\mathbf{x}, \mathbf{y}, t)$ and irreversibility $\Omega(\mathbf{x}, \mathbf{y})$ will be regarded as tensor fields.  For example, when one looks at correlations between different components of the displacements, e.g., $X_i(\mathbf{x}, t)$ and $X_j(\mathbf{x}, t)$ this naturally leads via similar steps as those above to definitions of tensor fields $a_{ij}(\mathbf{x}, \mathbf{y}, t)$ and $\Omega_{ij}(\mathbf{x}, \mathbf{y})$.  In the present paper we restrict ourselves to consider only one such component, i.e., the scalar field case.
 
\section{Elastic and fluctuational forces of the overdamped dynamics }
\label{sec:general model}

The overdamped dynamics is expressed by an instantaneous balance between viscous, elastic and fluctuation forces at each $\mathbf{x}$ in $R$, 
\begin{equation}
\partial_{t}X = \mu (-\mathcal{L}X + f).
\label{eq:Eq_motion_general}
\end{equation}
Here, $\mu$ is the \textit{mobility} associated with viscous drag, assumed to be independent of position and time.  The fluctuation force $f = f(\mathbf{x}, t)$ is a random variable whose statistics we describe shortly.  In general, a linear elastic restoring force is characterized by a linear operator $\mathcal{L}$ acting on the field $X(\mathbf{x}, t)$ in $R$, subject to boundary conditions on $\partial R$.  The \textit{elastic potential energy} $E$ of the filament is the quadratic form of $\mathcal{L}$, i.e.,
\begin{equation}
E = \frac{1}{2} \int_R X \; \mathcal{L}X \; \mathbf{dx}.
\end{equation}
 The operator together with the associated boundary conditions is \textit{symmetric} in the usual sense,
\begin{equation}
\int_R F \; \mathcal{L}G \; \mathbf{dx} = \int_R \mathcal{L}F \; G \; \mathbf{dx},  \nonumber
\end{equation}
for  all $F(\mathbf{x}), G(\mathbf{x})$ which satisfy the boundary conditions.  The operator $\mathcal{L}$ is positive, meaning $E > 0$ for all $X(\mathbf{x}, t)$ not identically zero satisfying the boundary conditions.   The elastic restoring force per unit length is \textit{minus} the variational derivative of potential energy,     
\begin{equation}
-\frac{\delta E}{\delta X} = -\mathcal{L}X.
\end{equation}

We can descend onto particular examples of continuum objects at will.  In the simplest \textit{string model},  the continuum is a tensioned string between fixed endpoints.   In this case, the region $R$ is the line segment of $x$ axis, $0 < x < L$, and fixed endpoints means zero boundary conditions  
\begin{equation}
X(0, t) = X(L, t) = 0.
\end{equation}
The elasticity operator $\mathcal{L}$ is
\begin{equation}
\mathcal{L} = -\kappa \partial_{xx},
\end{equation}
where $\kappa$ is the string tension.  We discern the positivity of $\mathcal{L}$ from the identity
\begin{equation}
\int_0^L X(-\kappa \partial_{xx}X)\; dx =  \kappa \int_0^L (\partial_x X)^2 \; dx,
\end{equation}
which follows from an integration by parts and use of the zero boundary conditions.

We work with dimensionless equations.  We can always choose units in which the dimensionless mobility $\mu$ is unity.  For instance, the scaling units
\begin{equation}
[X] = [x] = L, \: [t] = \frac{L}{\mu \kappa}, \: [f] = \frac{\kappa}{L}, \: [E] = \kappa L.
\end{equation}
applied to the string model lead to dimensionless equations  in which the mobility $\mu$, the string tension $\kappa$ and length $L$ are all unity.  


The statistics of the fluctuation force $f(\mathbf{x}, t)$ per unit length is assumed to be stationary.  Its time-independent mean value induces a corresponding time-independent mean displacement configuration of the filament.  In order to concentrate on \textit{fluctuations}, we consider the displacement field \textit{relative} to this mean configuration, and the effective force is the fluctuation of the total force about the mean.  Hence, there is no loss in generality by assuming that the mean force is identically zero.  Its two event correlation is expressed by
 \begin{equation}
\langle f(\mathbf{x}, t) f(\mathbf{x'}, t') \rangle  = 2 D(\mathbf{x}, \mathbf{x'}) \delta(t-t').
\end{equation}
The proportionality to $\delta(t-t')$ implies stationarity and negligibly short correlation time \footnote{We have also carried out calculations for the case of nonvanishing noise correlation times and they reveal behavior that is qualitatively similar to the principal results of this paper. This includes the form of the generalized flucutuation-dissipation relation and the behavior of variance mode plots, discussed in Sections V and VI.}.  Reversibility, or invariance under exchange of $t$ and $t'$ implies $D(\mathbf{x}, \mathbf{x'})$ is symmetric in $\mathbf{x}$ and $\mathbf{x'}$.  Although the force statistics is assumed here to be reversible, we see below that irreversible string statistics is nonetheless possible.

To discern the physical meaning of $D(\mathbf{x}, \mathbf{x'})$, we examine the ensemble of displacements $X(\mathbf{x}, t)$ driven \textit{only} by the force term $f(\mathbf{x}, t)$, such that
$\partial_t X =  f$ for $t > 0$ and subject to zero initial condition $X(\mathbf{x}, 0) = 0$.  An elementary calculation shows that
\begin{equation}
\langle X(\mathbf{x}, t) X(\mathbf{x'}, t) \rangle = 2 D(\mathbf{x}, \mathbf{x'}) \; t.
\end{equation}
The growth of the two point correlation proportional to elapsed time is traditionally called diffusion; hence, we call $D(\mathbf{x}, \mathbf{x'})$ the \textit{diffusion field}.

 Since the random force $f$ is the sum of statistically independent thermal and active components, the diffusion field can also be expressed as the sum of thermal and active components.  Below, we see that the part of the diffusion field, $D^{eq}$, which supports equilibrium string statistics at temperature $T$ satisfies an Einstein relation that can be expressed in the form
 \begin{equation}
 D^{eq}(\mathbf{x}, \mathbf{x'}) =  T \; \delta(\mathbf{x}-\mathbf{x'}).
 \end{equation}
However, only the active component induces irreversibility.  In the  biophysical picture of an active medium,  the molecular motors which are sources of active forces typically do not couple directly to the continuum probe \cite{MacKintosh_PRL_2008, Gladrow_PRL_2016}.  The motors and the probe are both embedded in a space-filling viscoelastic network which is the medium of communication between motors and the probe.  Thus, localized perturbations originating at a particular motor are typically somewhat spread out at the regions where they contact the probe.   A simple model of this spread out force is the product of a spatial \textit{footprint} $\phi(\mathbf{x})$ and a random function of time \cite{Brangwynne_PRL_2008, Gladrow_PRE_2017}.  The contribution to the diffusion from the motor is modeled by the product of normalized footprints,
\begin{equation}
D^{motor}(\mathbf{x}, \mathbf{x'}) = \phi(\mathbf{x}) \phi(\mathbf{x'}).
\end{equation}
The diffusion field due to several motors inducing statistically independent disturbances is a sum of products like (20).

\section{Amplitude statistics - general formulation}
\label{sec:Stat_ampl}

We denote the orthonormal eigenfunctions of the elasticity operator and associated boundary conditions by
\begin{equation}
u_n(\mathbf{x}), \: n = 1, 2, \dots .
\end{equation}
These satisfy
\begin{equation}
\mathcal{L} u_n(\mathbf{x}) = \lambda_n u_n(\mathbf{x})
\end{equation}
for $\mathbf{x}$ in $R$ subject to the appropriate boundary conditions on $\partial R$.  Here, $\lambda_n$ is the eigenvalue corresponding to eigenfunction $u_n(\mathbf{x})$ and must be positive because the operator $\mathcal{L}$ is positive.  
The eigenfunctions form a basis for representing scalar fields, denoted by $g(\mathbf{x})$, in $R$.  The coefficient of $u_n(\mathbf{x})$ in the expansion of $g(\mathbf{x})$ is denoted by $g_n$.   Analogously, the products
\begin{equation}
u_m(\mathbf{x}) u_n(\mathbf{y}) 
\end{equation}
constitute an orthonormal basis for scalar fields $g(\mathbf{x}, \mathbf{y})$ on $R \times R$.  We denote the coefficient of $u_m(\mathbf{x}) u_n(\mathbf{y})$ in the expansion of $g(\mathbf{x}, \mathbf{y})$ by $g_{mn}$.  We call the expansion coefficients $g_n$ or $g_{mn}$ the \textit{amplitudes} of the respective fields $g(\mathbf{x})$ or $g(\mathbf{x}, \mathbf{y})$.  For scalar fields $g(\mathbf{x})$ on $R$ we have the identity
\begin{equation}
(\mathcal{L} g)_n = \lambda_n g_n.
\end{equation}
For scalar fields on  $R \times R$,
\begin{equation}
(\mathcal{L}_x g)_{mn} = \lambda_m g_{mn}, \:\: (\mathcal{L}_y g)_{mn} = \lambda_n g_{mn}.
\end{equation}
Here, $\mathcal{L}_x g(x, y)$ denotes the application of the elasticity operator $\mathcal{L}$ on $g(\mathbf{x}, \mathbf{y})$ as a function of $\mathbf{x}$, with $\mathbf{y}$ fixed.  Similarly, $\mathcal{L}_y g(\mathbf{x}, \mathbf{y})$ means application of the operator $\mathcal{L}$ to $g(\mathbf{x}, \mathbf{y})$ as a function of $\mathbf{y}$ with $\mathbf{x}$ fixed. 

The dynamics (10) can be expressed in the amplitude representation as
\begin{equation}
\dot{X}_n = - \lambda_n X_n + f_n.
\end{equation}
For a given force amplitude $f_n(t)$, all solutions for the displacement amplitude $X_n(t)$ converge to the particular solution
\begin{equation}
X_n(t) = \int_0^\infty e^{-\lambda_n t'} f_n(t-t')\; dt',
\end{equation}
as $t \rightarrow \infty$.  This particular solution has no memory of initial conditions, and we call it the \textit{no memory solution}.  
Since the no memory solution maps each force amplitude $f_n(t)$ into a corresponding displacement amplitude $X_n(t)$, there is a unique mapping of force amplitude correlations into displacement amplitude correlations,
\begin{eqnarray}
 & & \langle X_m(t) X_n(t + \tau) \rangle =  \nonumber \\ & & \:\:\:\: \int_0^\infty \int_0^\infty e^{-\lambda_m t' - \lambda_n t''} \langle f_m(t') f_n(t''-\tau) \rangle  \; dt'' dt',  \nonumber
\end{eqnarray}
where we have used the property
\begin{equation}
\langle f_m(t-t') f_n(t+\tau-t'') \rangle =  \langle f_m(t') f_n(t''-\tau) \rangle,  \nonumber
\end{equation}
which follows from the stationarity and reversibility of force statistics.  The translation of the force correlation equation (17) into amplitude form can be written as
\begin{equation}
\langle f_m(t) f_n(t') \rangle = 2D_{mn} \delta(t-t'),  \nonumber
\end{equation}
where
\begin{equation}
D_{mn} := \int_{R \times R}d\mathbf{x} d\mathbf{y} D(\mathbf{x},\mathbf{y})u_m(\mathbf{x})u_n(\mathbf{y}).  \nonumber
\end{equation}
Hence
\begin{equation}
\langle X_m(t) X_n(t + \tau) \rangle =  2D_{mn} e^{-\lambda_n \tau}  \int_0^\infty e^{-(\lambda_m  + \lambda_n) t'} \; dt'   \nonumber
\end{equation}
or
\begin{equation}
\langle X_m(t) X_n(t + \tau) \rangle =   \frac{2 D_{mn}}{\lambda_m + \lambda_n} e^{-\lambda_n \tau} .  
\end{equation}

We read off the basic equations of field statistics in the mode amplitude formulation.  Setting the time lag $\tau$ in (28) to zero, we have
\begin{equation}
M_{mn} := \langle X_m(t) X_n(t) \rangle =  \frac{2D_{mn}}{\lambda_m + \lambda_n}.  
\end{equation}
Following traditional nomenclature, this mapping of diffusion amplitudes $D_{mn}$ into second moments $M_{mn}$ of displacement amplitudes can be viewed as a generalized fluctuation-dissipation relation. 

Next, we show how a diffusion field $D^{eq}(\mathbf{x}, \mathbf{y})$ can be determined for the filament driven by equilibrium thermal fluctuations at temperature $T$.   The elastic energy (11) in terms of normal mode amplitudes is
\begin{equation}
E = \frac{1}{2} \sum_n \lambda_n X_n^2. \nonumber
\end{equation}
At equilibrium, the marginal probability density $\rho^{eq}$ of the $N$ amplitudes $X_1, \dots X_N$ is proportional to the Boltzmann factor,
\begin{equation}
\rho^{eq} \propto e^{-\frac{1}{2T} \sum _1^N \lambda_nX_n^2}.  \nonumber
\end{equation}
We read off its second moments
\begin{equation}
M_{mn} = \langle X_m X_n \rangle = \frac{T}{\lambda_m} \; \delta_{mn}.  \nonumber
\end{equation}
The amplitudes of the diffusion field now follow from the fluctuation dissipation relation (29),
\begin{equation}
(D^{eq})_{mn} =  T \; \delta_{mn}.  \nonumber
\end{equation}
Thus, the equilibrium diffusion field can be expressed as
\begin{eqnarray}
& & D^{eq}(\mathbf{x}, \mathbf{y}) = \sum_{m, n} (D^{eq})_{mn} u_m(\mathbf{x}) u_n(\mathbf{y}) = \nonumber \\  & & \:\:\:\:\:\:\:\:\:\:\:\:\:\:\:\:\:\:\:\:\:\:\:\ \:\:\:\:\:\:\:\:\:\:\:\: T \sum_m u_m(\mathbf{x}) u_m(\mathbf{y}),  \nonumber
\end{eqnarray}
or finally, using the completeness of the eigenfunctions $\{u_n\}$,
\begin{equation}
D^{eq}(x, y) = T \; \delta(\mathbf{x}-\mathbf{y}).  
\end{equation}
The classical equilibrium statistics presented here has a fundamental limitation.  The expected potential energy in each degree of freedom $X_n$ is $T/2$.  Since the degrees of freedom $X_1, X_2, \dots$ of a \textit{continuous} filament are countably \textit{infinite}, the equilibrium thermal energy diverges - a well-known defect in the classical thermodynamics of continuous fields.  In practice, a real filament with its discrete atomic structure has a finite number of degrees of freedom.  Here, we concentrate on features of the filament statistics that are finite in the continuum approximation.

We now show how the irreversibility field amplitudes can be related to empirical maps of mode amplitude correlations of the type that have been measured in recent biophysics experiments \cite{Bacanu_2023}.  The amplitudes of the irreversibility field with time lag $\tau$, cf. (3), can be written as
\begin{equation}
\omega_{mn}(\tau) := \frac{1}{2} \langle X_m(t+\tau)X_n(t) - X_m(t) X_n(t+\tau) \rangle.
\end{equation}
Then, by (28), we have the explicit formula
\begin{equation}
\omega_{mn}(\tau) = \frac{D_{mn}}{\lambda_m + \lambda_n} \left( e^{-\lambda_m \tau} - e^{-\lambda_n \tau} \right). 
\end{equation}
 In the limit $\tau \rightarrow 0^+$, we recover the corresponding amplitudes of the irreversibility field defined by (4),
\begin{equation}
\Omega_{mn}  = \frac{\lambda_n - \lambda_m}{\lambda_m + \lambda_n} D_{mn}.  
\end{equation}
The amplitudes of the irreversibility field have direct, physical meaning.  From the translations of (7) and (9) into amplitudes, we deduce
\begin{equation}
\lim_{t \rightarrow \infty} \frac{1}{2t} \int_0^t (X_m \dot{X}_n -  \dot{X}_mX_n)(t') \; dt' = -\Omega_{mn}.
\end{equation}
On the left hand side we recognize the stochastic area in the plane of the $m$-th and $n$-th mode amplitudes.  If we compute this stochastic area using experimentally recorded time series of these amplitudes, then according to (9), the empirical stochastic area divided by $t$ converges to $-\Omega_{mn}$ as $t \rightarrow \infty$.

\section{Amplitude statistics - the string case}
\label{sec:Stat_ampl_string}

In this section, we illustrate the general results of the preceding sections with a concrete example: we compute irreversibility amplitudes of a string subject to the active force of a single motor.  The string eigenvalues and eigenfunctions are 
 \begin{equation}
 \lambda_n = \pi^2 n^2, \:\: u_n(x) = \sqrt{2} \sin n \pi x, \: 0 < x < 1.
 \end{equation}
The diffusion field of a motor with footprint $\phi(x)$ is given by (20), whose translation into amplitudes is 
\begin{equation}
D_{mn} = \phi_m \phi_n.
\end{equation}
Substituting (35) and (36) into (33), we have
\begin{equation}
\Omega_{mn} = \frac{n^2 - m^2}{m^2 + n^2} \; \phi_m \phi_n.
\end{equation}

We take the footprint to be a Gaussian centered about position $x = a$ along the string, with $0 < a < 1$, and given by 
 \begin{equation}
\phi(x) = \sqrt{\frac{d}{2 \pi \sigma}} e^{-\frac{(x-a)^2}{2 \sigma}}. 
\end{equation}
This footprint has characteristic length $\sqrt{\sigma}$.  The constant $d$ represents the \textit{strength} of the motor, so that its diffusion field satisfies 
\begin{equation}
\int_{\mathbb{R}^2} D(x, y) \; dx dy = d.  \nonumber
\end{equation}
 Its amplitudes are
\begin{eqnarray}
 & & \phi_m = \int_0^1 \phi(x) u_m(x) \; dx =  \nonumber \\ & & \:\:\:\:\: \:\:\:\:\: \sqrt{\frac{d}{2 \pi \sigma}} \int_0^1 e^{-\frac{(x-a)^2}{2 \sigma}} \sin m \pi x \; dx.  \nonumber
\end{eqnarray}
Asymptotic evaluation of the integral in the limit $\sigma << 1$ gives
\begin{equation}
\phi_m \sim \sqrt{2d} \; e^{-\frac{\pi^2 \sigma}{2} m^2}  \; \sin m \pi a.
\end{equation}

\begin{figure}[!h]
\centerline{\includegraphics[width=.50\textwidth]{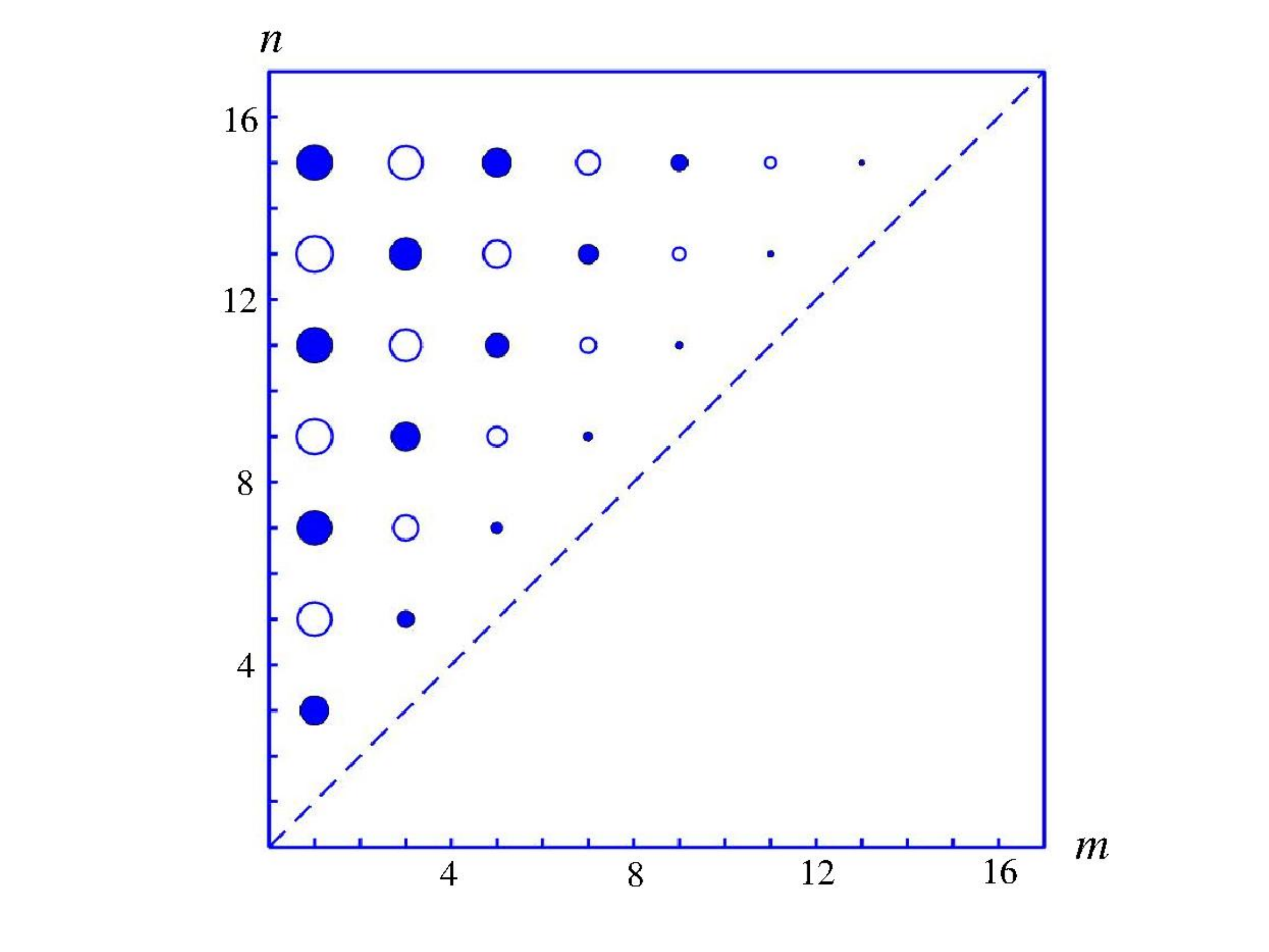}}
\caption{Irreversibility field amplitudes $\Omega_{nm}$ for a single active source with delta-function footprint, i.e., $\sigma = 0$ placed symmetrically at $a = 0.50$, i.e., the center of the filament.}
\label{FIG2}
\end{figure}

\begin{figure}[!h]
\centerline{\includegraphics[width=.50\textwidth]{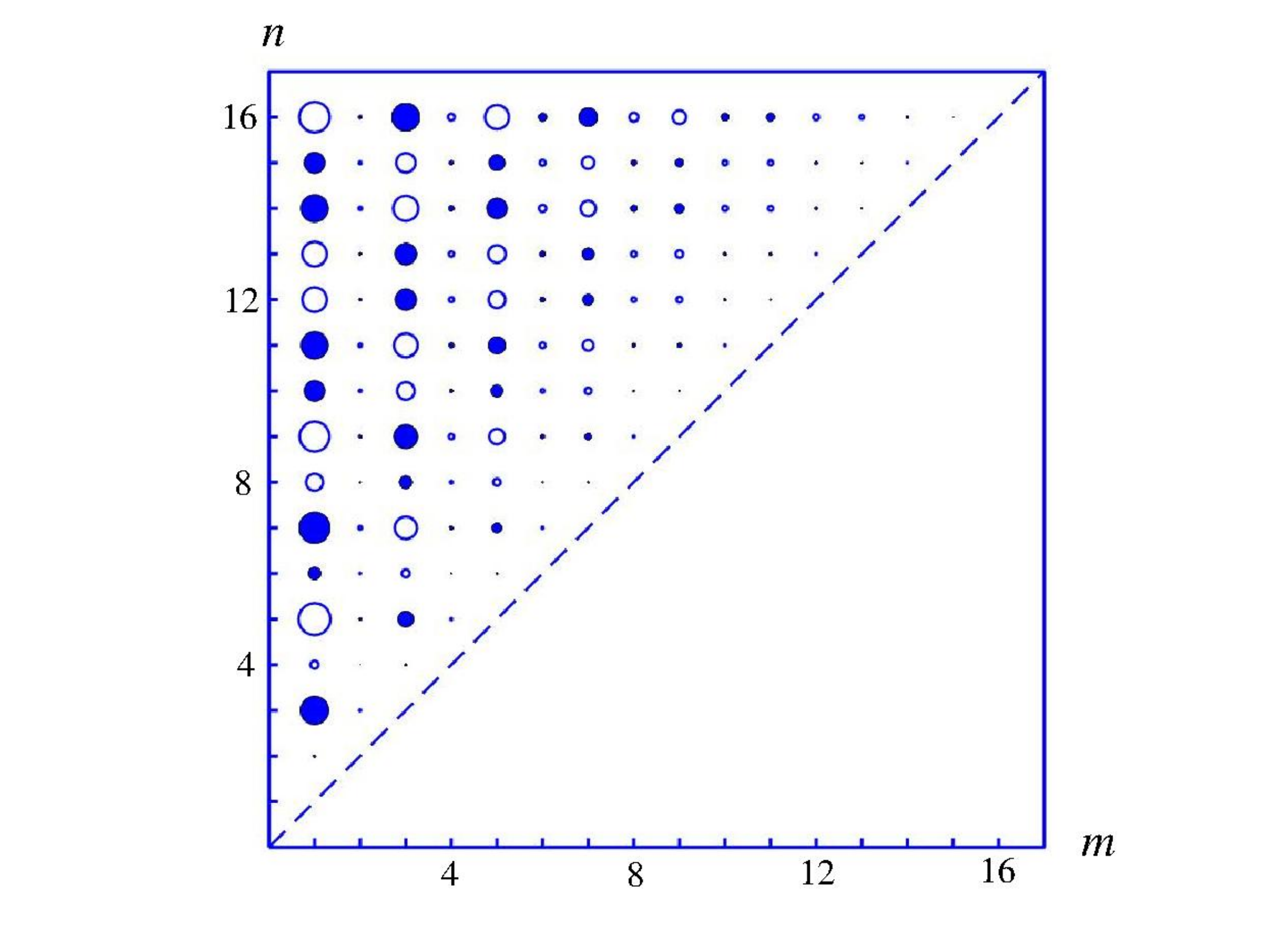}}
\caption{Irreversibility field amplitudes $\Omega_{nm}$ for a single active source with delta-function footprint, i.e., $\sigma = 0$ placed at $a = 0.52$, i.e., displaced slightly from the center of the filament.}
\label{FIG3}
\end{figure}

We construct a graphical representation of $\Omega_{mn}$ as a function of integer lattice points $(m, n)$.  Due to the antisymmetry in $m$ and $n$, it is sufficient to examine the restriction to the triangular sub-lattice with $n > m$.  About each lattice point 
$(m, n)$ we construct a disk whose radius is proportional  to $|\Omega_{mn}|$.   The disk is filled for $\Omega_{mn}$ positive,
and unfilled if negative.  Let's take the simplest case $\sigma = 0$, i.e., the footprint of the motor is a single point.  Figures \ref{FIG2} and \ref{FIG3} are the graphical representations of $\Omega_{mn}$ for motor positions $a = 0.50$ and $0.52$ respectively.  We see that a very small change in motor position induces a large, qualitative change in the irreversibility field $\Omega_{mn}$.  This is not surprising since $\Omega_{mn}$ contains the factor $\sin m \pi a \sin n \pi a$ whose $a$ dependence is highly oscillatory for mode numbers $m, n >> 1$.

\begin{figure}[!h]
\centerline{\includegraphics[width=.50\textwidth]{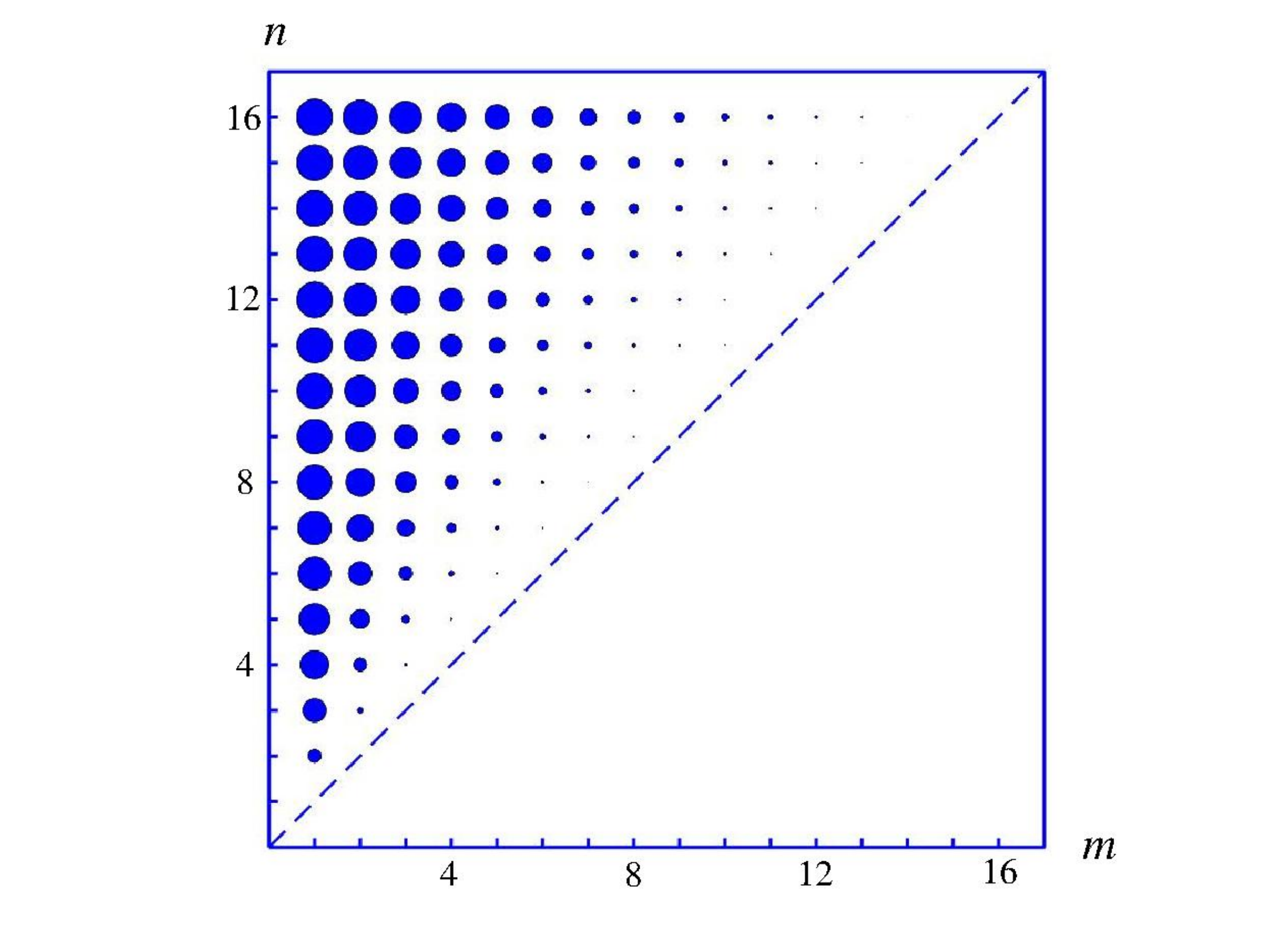}}
\caption{Irreversibility field variances $\langle \Omega_{nm}^2 \rangle$ for a single active source with delta-function footprint, i.e., $\sigma = 0$ placed symmetrically at $a = 0.50$.}
\label{FIG4}
\end{figure}

For biophysical systems such as filaments embedded in active viscoelastic networks, there is typically little control over motor positions \cite{Bacanu_2023, Gradziuk_2019, Gladrow_PRE_2017}.  Given the sensitive dependence on $a$ it makes sense to consider the motor position $a$ as a random variable, and $\Omega_{mn}$ as a dependent random variable.  If the random motor position has a uniform distribution on $0 < a < 1$, the mean of $\Omega_{mn}$ is clearly zero.  However, the variance is nonzero and given by
\begin{equation}
\langle\Omega_{mn}^2 \rangle =  d^2 \left( \frac{n^2 - m^2}{m^2 + n^2} \right)^2 \; e^{-\pi^2 \sigma (m^2 + n^2)}.
\end{equation}
Figure \ref{FIG4} is the graphical representation of the variance restricted to the triangular sub-lattice for $\sigma = 0$, i.e., a delta-function footprint.  The radius of the disk about any lattice point is proportional to the variance at that point.   

\begin{figure}[!h]
\centerline{\includegraphics[width=.50\textwidth]{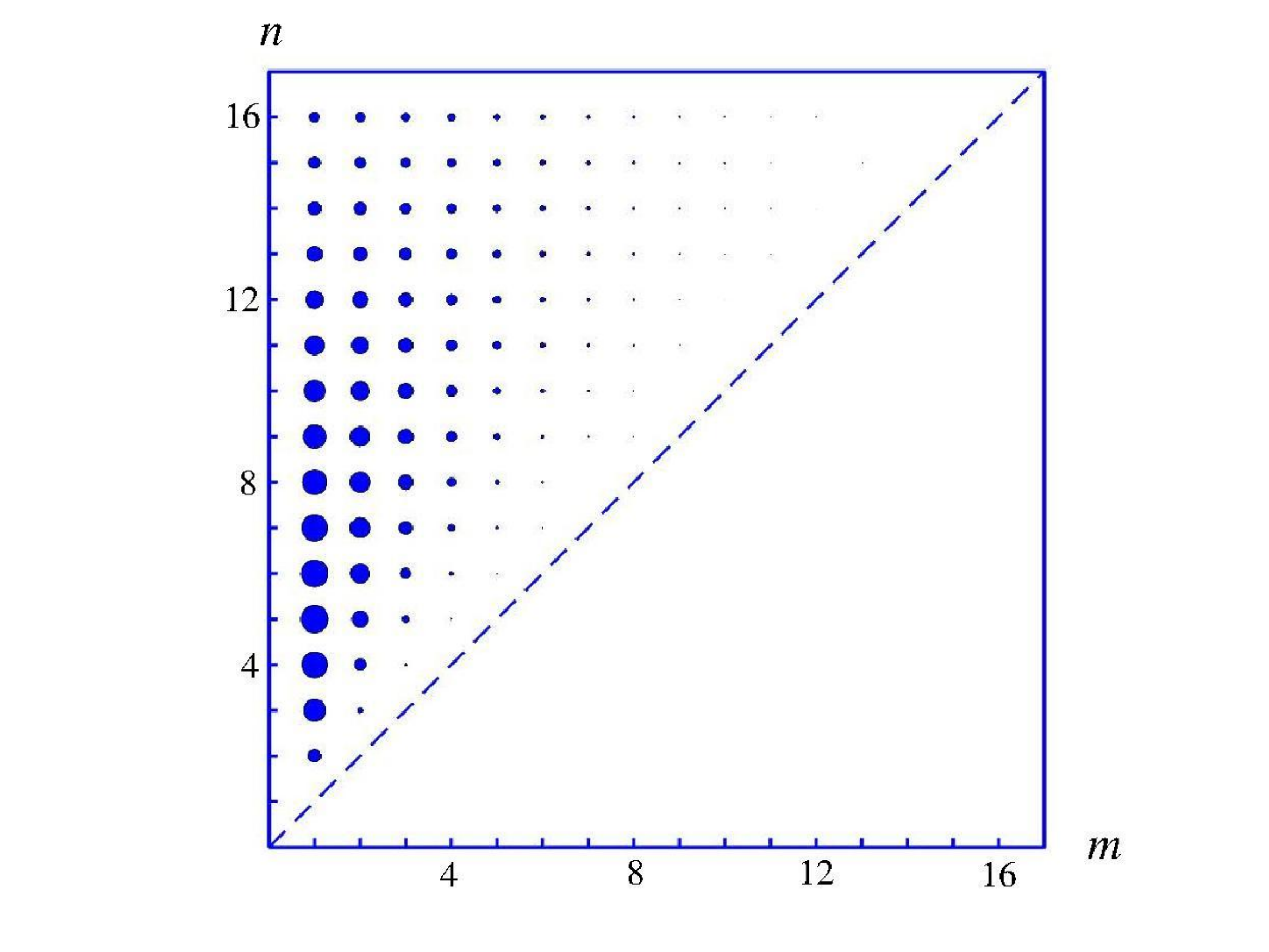 }}
\caption{Irreversibility field variance $\langle \Omega_{nm}^2 \rangle$ for a single active source with non-zero footprint ($\sigma = 0.0005$) and placed symmetrically at $a = 0.50$, i.e., the filament center.}
\label{FIG5}
\end{figure}

To see the effect of \textit{finite} motor footprint, we set $\sigma = 0.0005$ in Fig. \ref{FIG5}, so that the footprint has characteristic length $\sqrt{0.0005} \approx 0.0223$, which is about $2\%$ of the total string length.  The effect of the exponential term in (40) is immediately evident and causes strong attenuation of amplitudes for larger wavenumbers, i.e., those with small characteristic lengths.  Overall, the features of Fig. \ref{FIG5} are qualitatively similar to those found in the recent experimental data of Bacanu \textit{et al.} \cite{Bacanu_2023}.  One striking difference concerns the behavior of the variances as one approaches the diagonal.  In the experiment, there is no apparent decrease as $m \rightarrow n$; however, the analytical results show a clear diminution which is apparent from (40), from the factor that scales like $(m^2 - n^2)^2$.  Qualitatively similar results \cite{Neu_preprint_2024} are found analytically when the string is replaced by a rod with stiffness which models an embedded nanotube more accurately, so that is not the reason for the discrepancy.  We conjecture effects that could allow for closer crowding of variance as the diagonal is approached include mode coupling effects or nonlinear effects.


\medskip
  
\section{The spatial formulation of string statistics}
\label{sec:Space_stat} 

The spatial formulation of stationary continuum statistics examines the irreversibility field $\Omega(\mathbf{x}, \mathbf{y})$ defined in (4)  as a function of $(\mathbf{x}, \mathbf{y})$ in  $R \times R$.  The clearest presentation begins with the \textit{displacement correlation field}
\begin{equation}
M(\mathbf{x}, \mathbf{y}) := \langle X(\mathbf{x}, t) X(\mathbf{y}, t) \rangle,
\end{equation}
and the determination of the irreversibility field from it.  First, we translate the generalized fluctuation-dissipation relation (29) into its spatial formulation.  We start by writing (29) as
\begin{equation}
(\lambda_m + \lambda_n) M_{mn} = 2 D_{mn}. \nonumber
\end{equation}
Then, using the identities in (25), we have
\begin{equation}
((\mathcal{L}_x + \mathcal{L}_y)M)_{mn} = 2 (D)_{mn} \nonumber
\end{equation}
for all mode number pairs $m$ and $n$.  Hence, we arrive at the continuum form of the fluctuation-dissipation relation,
\begin{equation}
(\mathcal{L}_x + \mathcal{L}_y)M = 2D,
\end{equation}
for all $(\mathbf{x}, \mathbf{y})$ in $R \times R$.  

For the string elasticity operator $\mathcal{L}$ as in (14), (42) takes the explicit form of a Poisson equation
\begin{equation}
(\partial_{xx} + \partial_{yy}) M = -2D
\end{equation}
for $(x, y)$ in the square $S$ defined by $0 < x, y < 1$.  The zero endpoint conditions on the displacement field $X(x, t)$ at $x = 0, 1$ imply a zero boundary condition for $M$:
\begin{equation}
M \equiv 0, \: (x, y) \in \partial S.
\end{equation}
In summary, the spatial form of the generalized fluctuation-dissipation relation is a linear inhomogeneous boundary value problem for $M$ which has \textit{twice} the dimension of the physical object that is undergoing fluctuations, and for which the diffusion field $D$ acts as the source of $M$.

In order to determine the irreversibility field $\Omega$ from $M$, we rewrite (33) for the irreversibility amplitudes $\Omega_{mn}$ in the following form
\begin{equation}
(\lambda_m + \lambda_n) \Omega_{mn} =  (\lambda_n - \lambda_m) D_{mn}.  \nonumber
\end{equation}
Evoking the identities expressed by (25) once more, we write
\begin{equation}
(\mathcal{L}_x + \mathcal{L}_y) \Omega = (\mathcal{L}_y - \mathcal{L}_x) D. \nonumber
\end{equation}
Alternatively, apply the operator $\mathcal{L}_y - \mathcal{L}_x$ to the continuum form of the fluctuation-dissipation relation (42) to find
\begin{equation}
(\mathcal{L}_x + \mathcal{L}_y) (\mathcal{L}_y - \mathcal{L}_x) M = 2(\mathcal{L}_y - \mathcal{L}_x) D.  \nonumber
\end{equation}
Due to the positivity of the operator $\mathcal{L}_x + \mathcal{L}_y$, comparison of the last two equations gives the unique identification for irreversibility field expressed in terms of displacement correlation field, 
\begin{equation}
\Omega = \frac{1}{2}(\mathcal{L}_y - \mathcal{L}_x) M.
\end{equation}
For the special case of the string, we have
\begin{equation}
\Omega(x, y) = \frac{1}{2} (\partial_{xx} - \partial_{yy}) M.
\end{equation}
The spatial formulation of continuum statistics consists of the boundary value problem for $M$ and the determination of $\Omega$ from $M$.  It allows us to see where irreversibility occurs in space.
  
\begin{figure}[!h]
\centerline{\includegraphics[width=.50\textwidth]{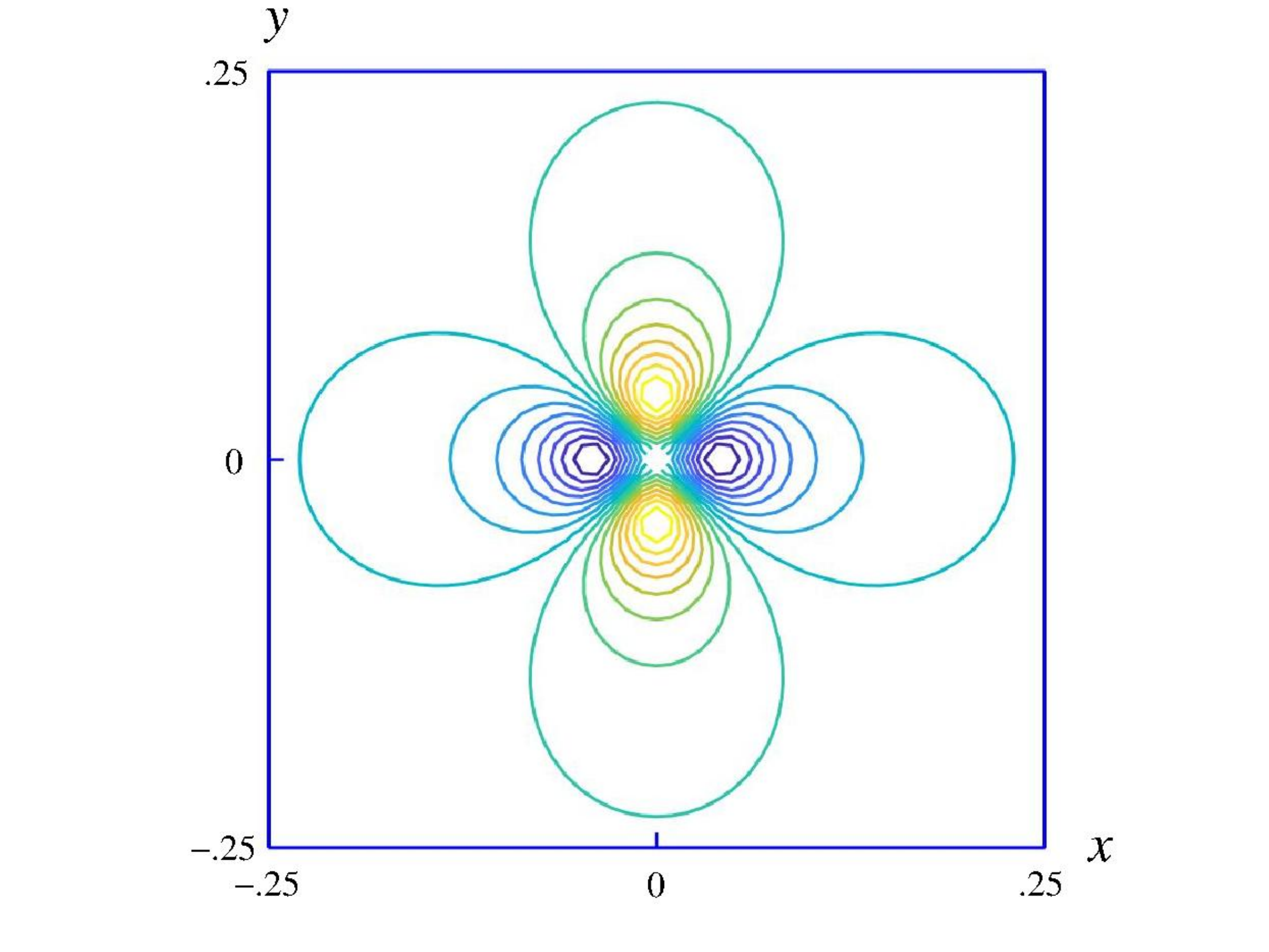}}
\caption{Quadrupole structure of irreversibility field for the string with a single active noise source placed at the origin with footprint $\sigma = 0.0005$.}
\label{FIG6}
\end{figure}

\begin{figure}[!h]
\centerline{\includegraphics[width=.50\textwidth]{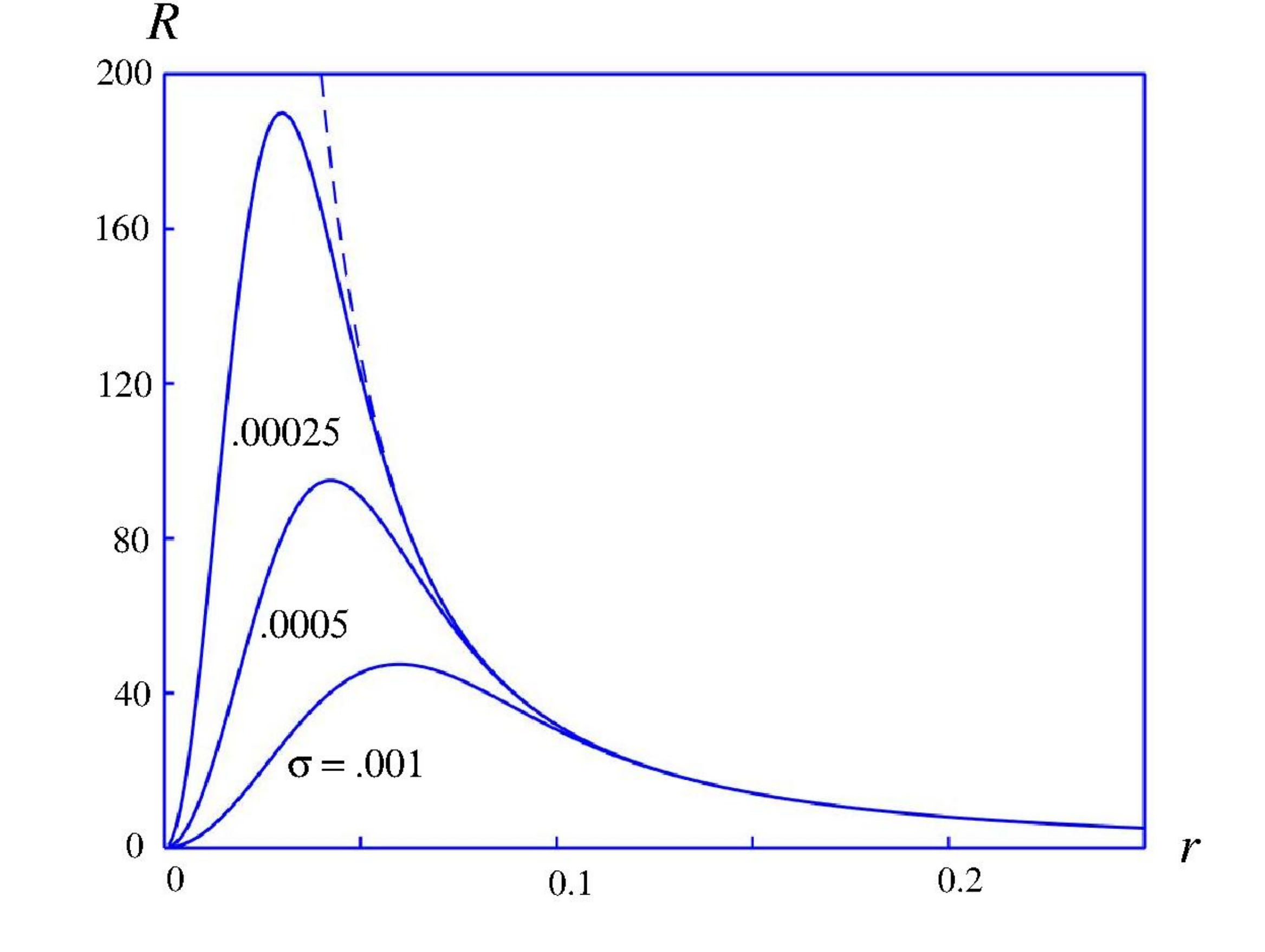}}
\caption{Radial factor $R$ as a function of $r$ for different values of motor footprint parameter $\sigma = 0.00025, \; 0.0005, \; 0.001$. Dashed curve shows the (diverging) radial factor for a delta source,  $\sigma = 0.$}
\label{FIG7}
\end{figure}

For instance, let's examine the correlation and irreversibility fields of a string subject to a motor with the Gaussian footprint as in (38).   We shift the origin of $x$ so $a = 0$, and the diffusion field is 
\begin{equation}
D(x, y) = \frac{d}{2 \pi  \sigma}  e^{-\frac{x^2 + y^2}{2 \sigma}}.  \nonumber
\end{equation}
The Gaussian footprint implies that $D$ has circular symmetry about the origin of the $x, y$ plane.  Furthermore, in the limit $r \rightarrow 0$, $M$ is also expected to be circularly symmetric so that the Poisson equation (43) reduces to
\begin{equation}
M'' + \frac{1}{r} M'  = -\frac{d}{ \pi \sigma} \; e^{-\frac{r^2}{2 \sigma}}.
\end{equation}
The solution can be written as
\begin{equation}
M'(r) =  -\frac{d}{ \pi} \frac{1 - e^{-\frac{r^2}{2 \sigma}}}{r},
\end{equation}
where we have imposed a boundary condition that $M'$ is regular at $r = 0$.
We recover the irreversibility field from 
\begin{equation}
\Omega = \frac{1}{2} (\partial_{xx} - \partial_{yy} ) M =  R(r) \; \frac{x^2-y^2}{r^2}.
\end{equation}
Here, 
\begin{eqnarray}
& & R(r) := \frac{1}{2} \; \left( M'' - \frac{M'}{r} \right) = \nonumber \\ & & \:\:\:\:\:\:\:\:\:\:\:\: \frac{d}{2 \pi r^2} \left(1  - \left(1 + \zeta \right) e^{-\zeta} \right), \; \zeta := \frac{r^2}{2 \sigma}.
\end{eqnarray}
The asymptotic behavior as ${r}/\sqrt{\sigma} \rightarrow \infty$ is given by
\begin{equation}
\Omega \sim \frac{d}{2 \pi} \; \frac{x^2-y^2}{r^4},
\end{equation}
a two-dimensional quadrupole field.   This is the solution we  would get for the point source diffusion field $D(x, y) = d \; \delta(x) \delta(y)$.  Figure \ref{FIG6} depicts the contours of $\Omega (x, y)$ in the $x, y$ plane for $\sigma = 0.0005$ and shows clear features of quadropule field for large $r$.  Figure \ref{FIG7} depicts the radial factor $R(r)$ for different values of $\sigma$ and shows how this function approaches the diverging form that applies for zero footprint.  It should be noted that the radial positions of the local maxima scale like $\sigma^{1/2}$ which follows from (50).

Recent studies of noise-driven discrete elastic network models \cite{Gradziuk_2019, Mura_2019} have examined how two point measures of irreversibility scale with distances between pairs of nodes.  They have found that the exponents in these scaling laws depend on the dimension of the network as defined by the topology of its connections.  Here, we understand these scaling insights from a continuum model perspective.  Our metrics are area enclosing rates encoded in the irreversibility field , cf. (4).  Unlike cycling frequencies, area enclosing rates depend only on the active forces and \textit{not} on the thermal background. In analogy with previous work \cite{Gradziuk_2019, Mura_2019}, we take the elasticity operator to be minus the appropriate Laplacian.  A basic insight from the previous work is naturally presented in the context of a continuum formulation, namely, the specific from of the generalized fluctuation-dissipation relation (42),
\begin{equation}
(\triangle_x + \triangle_y) M = - 2D,
\end{equation}
is a Poisson equation for $(\mathbf{x}, \mathbf{y}) \in R \times R$, which has \textit{twice} the dimension of the physical continuum object.  

This means that the correlation and irreversibility fields for two or three dimensional continuums display geometric attenuations characteristic of four or six dimensions, respectively.  If we assume that the continuum is subject to a motor with a Gaussian footprint about the origin, then the diffusion field is
\begin{equation}
D(\mathbf{x}, \mathbf{y}) = \frac{d}{(2 \pi \sigma )^{\frac{N}{2}}} e^{-\frac{r^2}{2 \sigma}},
\end{equation}
where $N$ is the dimension of the continuum and $r$ is the radial variable of $\mathbb{R}^{2N}$, i.e., $r^2 =  |\mathbf{x}|^2 + |\mathbf{y}|^2$.  The strength $d$ of the motor is the integral of $D(\mathbf{x}, \mathbf{y})$ over $\mathbb{R}^{2N}$.  As in the analysis of the string, we ignore the effect of the boundary $\partial (R \times R)$ for $(\mathbf{x}, \mathbf{y})$ near the origin, and the correlation field is approximated as radially symmetric,
$M \approx M(r)$.  It satisfies
\begin{equation}
M'' + \frac{2N-1}{r} M' =  \frac{2 d}{(2 \pi \sigma )^{\frac{N}{2}}} e^{-\frac{r^2}{2 \sigma}}.
\end{equation}
The left hand side is the radially symmetric Laplacian in $2N$ dimensions.  Given the solution with $M'(r)$ regular at $r = 0$, we can compute the irreversibility field from (45) with $\mathcal{L} = \triangle$.  For clarity we consider its restrictions to two dimensional planes in $2N$ dimensional space.  Their parametric representations are
\begin{equation}
(\mathbf{x}, \mathbf{y}) = (x \hat{\mathbf{x}}, y  \hat{\mathbf{y}}),
\end{equation}
where $\hat{\mathbf{x}}, \; \hat{\mathbf{y}}$ are fixed unit vectors in $\mathbb{R}^N$ and $x, y$ are any real numbers.  The irreversibility field in this plane is given by (49), but the radial factor $R(r)$ is modified.   For a two dimensional continuum, we have
\begin{equation}
R(r) = \frac{2d}{\pi ^2 r^4} \left( 1 - \left( 1 + \zeta + \frac{\zeta^2}{2} \right) e^{-\zeta} \right),
\end{equation} 
where $\zeta := r^2/2 \sigma$ as before.  The contours of $\Omega$ in such a plane display a quadrupole structure qualitatively similar to Fig. \ref{FIG6} for the string case.  In the limit ${r}/\sqrt{\sigma} \rightarrow \infty$, $\Omega$ has the asymptotic behavior
\begin{equation}
\Omega \sim \frac{2d}{\pi^2} \; \frac{y^2 - x^2}{r^6}.  \nonumber
\end{equation}
In a fixed direction of four dimensional space, this asymptotic behavior scales as $r^{-4}$.   For a three dimensional continuum, the radial factor is
\begin{equation}
R(r) =  \frac{12d}{\pi^3 r^6}  \left( 1 - \left( 1 + \zeta + \frac{\zeta^2}{2} + \frac{\zeta^3}{6} \right) e^{-\zeta} \right).
\end{equation} 
As $r \rightarrow \infty$, $\Omega$ scales like $r^{-6}$, the asymptotic behavior of a quadrupole in six dimensions.  

We now show how the irreversibility field can be used to calculate and visualize local patterns of dissipation and work induced by the active forces.  The time rate of change of the elastic energy (11) is
\begin{eqnarray}
& & \dot{E} = \frac{1}{2} \int_R (\partial_t X \; \mathcal{L}X + X \; \mathcal{L}\partial_t X ) \; \mathbf{dx} =   \nonumber \\ & & \:\:\:\:\:\:\:\:\:\:\:\:\:\:\:\:\:\:\:\:\:\:\:\:\:\:\:\:\:\:\:\:\:\:\:\:\:\:\:\:\:\:\:\:\:\:\:\:\:\:\:  \int_R \mathcal{L}X \; \partial_t X \; \mathbf{dx},
\end{eqnarray}
where the second equality follows from the symmetry of $\mathcal{L}$ and associated boundary conditions.  Evoking the filament dynamical equation (10), we can write
\begin{equation}
\dot{E} = - \int_R (\partial_t X)^2 \; \mathbf{dx} + \int_R f \; \partial_tX \; \mathbf{dx}.
\end{equation}
Here, we recognize that
\begin{equation}
\Dot{q} :=  (\partial_t X)^2
\end{equation}
is the rate of dissipation per unit length, and
\begin{equation}
\Dot{w} := f \; \partial_t X
\end{equation}
is the rate of work per unit length done by the fluctuation forces $f(\mathbf{x}, t)$.  In stationary statistics, the ensemble averages of total dissipation and work balance,
\begin{equation}
\int_R \langle \Dot{q} - \Dot{w} \rangle  \; \mathbf{dx} = 0.
\end{equation}
This overall balance holds for both reversible and irreversible statistics.  Irreversibility is evident when we examine the  difference \begin{equation}
\langle \Dot{q} - \Dot{w} \rangle = \langle \left( \partial_t X - f \right) \partial_t X \rangle. 
\label{eq:pointwise_formula_1}
\end{equation}
\textit{point wise}.
 A calculation in Appendix A allows to express this difference in terms of the irreversibility field,
\begin{equation}
\langle \Dot{q} - \Dot{w} \rangle =  \frac{1}{2} (\mathcal{L}_x - \mathcal{L}_y) \Omega(\mathbf{x}, \mathbf{x}).
\label{eq:pointwise_formula_2}
\end{equation}
For reversible statistics with $\Omega \equiv 0$, the expected dissipation and work per unit length balance point wise.  Nonzero point wise difference between $\langle \Dot{q} \rangle$ and $\langle  \Dot{w} \rangle$ indicates irreversibility.  

We examine the pointwise difference between rates of dissipation and work along a string subject to a motor with a Gaussian footprint.   Substituting (49) for $\Omega$, (64) becomes. 
\begin{equation}
\langle \Dot{q} - \Dot{w} \rangle = -2 \left(\frac{R}{r} \right)' (r = \sqrt{2} |x|) .
\end{equation}
The solid curve Fig. 8 is the graph of $\langle \Dot{q} \rangle - \langle \Dot{w} \rangle$ versus $x > 0$ based on (64).
\begin{figure}[!h]
\centerline{\includegraphics[width=.50\textwidth]{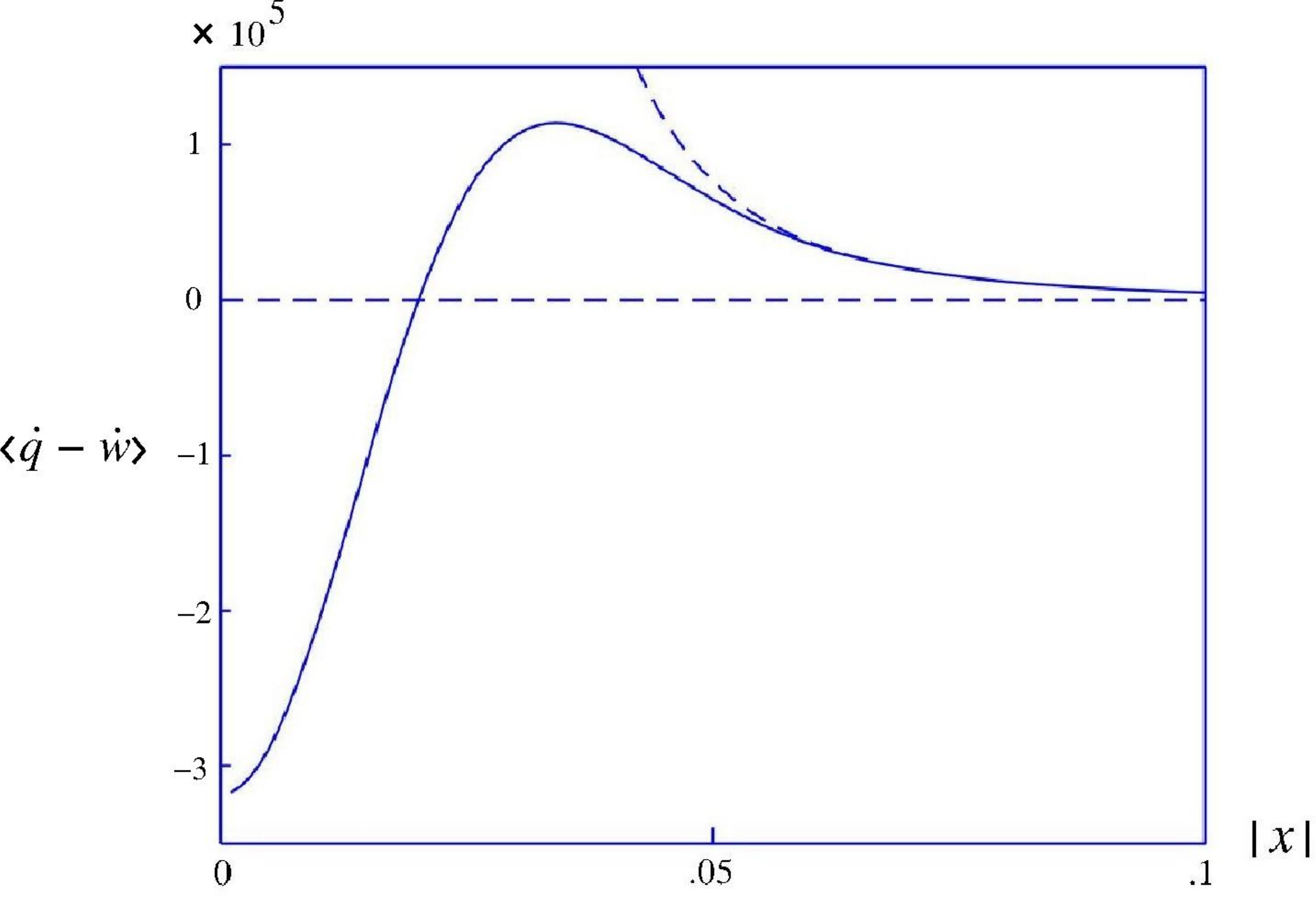}}
\caption{Point wise difference between work and dissipation vs. distance from the active source position for $\sigma = 0.001$.  }
\end{figure}
This graph is consistent with a simple intuition: the motor delivers power only to the portion of string which feels its force, within a characteristic length $\sqrt{\sigma}$ of the origin.  The deflection of the string extends beyond the support of the motor force, so the characteristic length of the dissipation is larger than the support of the force.  Hence, the negative values within characteristic distance
$\sqrt{\sigma}$ of the origin where the power from the motor is concentrated, and positive values farther away, where dissipation dominates.
The dashed curve in Fig. 8 shows the response in the limit $\sigma \rightarrow 0$ corresponding to a pointlike motor footprint and, in this case, the response is proportional to the inverse fourth power of $x$ as $x\rightarrow 0$.  The negative contribution near the origin due to the motor's power is not resolved at all.  On the other hand, the finite footprint of the motor regularizes this singular behavior at a characteristic lengthscale $O(\sqrt{\sigma})$.  Thus, an order of magnitude estimate for $\langle \Dot{q} - \Dot{w} \rangle$ is given by  $\sigma^{-2} = 10^6$ for $\sigma = 0.001$.  Hence the large scale of vertical axis in Fig. 8.

\medskip
 

\section{Concluding remarks}
\label{sec: Conclusions}

A general framework has been presented for quantitatively characterizing irreversibility effects in a class of spatially extended systems - namely linear viscoelastic systems driven by active and thermal noise sources with potential applications that include describing the fluctuation properties of various biophysical systems.  For such systems, we have extended the concept of stochastic area and have defined a closely related irreversibility field which plays a central role in understanding the relationship of fluctuational work done by active pointlike sources and the responding dissipative patterns that occur at larger length scales.  A generalized fluctuation-dissipation relation has been derived which takes the form of an elliptic boundary value problem - defined on a space with twice the dimension of the physical space of the filament - and driven by a diffusion field that is derived from the active noise sources acting on the filament.  Under steady state, non-equilibrium conditions, a general formula has been presented for calculating the irreversibility field from directly from the the two-point displacement correlation field which, in turn, is determined as the solution to the generalized flucutation-dissipation relation.  

The developed methods have been applied to the particular case where the embedded filament is an elastic string.  The results illustrate many of the qualitative behaviors of irreversibility fields expected for general linear elastic continuuum objects and they have the added advantage that exact analytical results can be obtained.  In separate work, we have studied properties of the irreversibility field for elastic \textit{rods} embedded in active media \cite{Neu_preprint_2024} which should provide, for example, a more accurate description of experiments that study carbon nanotubes embedded in living cells \cite{Bacanu_2023}.  In that case, we find qualitatively similar behavior in plots of the mode variances of the irreversibilty field (cf. Fig. 5) as well as in plots of the pointwise difference between work and dissipation (cf. Fig. 8).  On the other hand, since the rod elasticity operator includes a term proportional the fourth derivative of position (i.e., $\partial_{xxxx}$), the lefthand side of the generalized fluctuation-dissipation relation no longer possesses spherical symmetry and this leads to anisoptropy effects that are the focus of ongoing work.


\medskip


\section{Acknowledgements}
\label{sec: Acknowledgements}
 We thank Konstantin Matveev for his critical reading of on an early version of the manuscript.
 \bigskip

\appendix
\section{Derivation of expression for pointwise difference between fluctuational work and dissipation}
\label{sec:appendixA}

Here we derive the formula (\ref{eq:pointwise_formula_2}) for the point wise difference between local rates of dissipation and work.  Substituting for $\partial_tX$ from the dynamical equation (\ref{eq:Eq_motion_general}), (\ref{eq:pointwise_formula_1}) becomes
\begin{equation}
\langle \Dot{q} - \Dot{w} \rangle = \langle (\mathcal{L} X)^2 \rangle - \langle f \mathcal{L} X \rangle.
\end{equation}
Observe that
\begin{equation}
\langle \mathcal{L}_x X(\mathbf{x}, t) \; \mathcal{L}_y X(\mathbf{y}, t) \rangle = \mathcal{L}_x\mathcal{L}_y M(\mathbf{x}, \mathbf{y}),  \nonumber
\end{equation}
so that
\begin{equation}
\mu \langle (\mathcal{L} X (\mathbf{x}, t))^2 \rangle  = \mu \mathcal{L}_x \mathcal{L}_y M(\mathbf{x}, \mathbf{x}).
\end{equation}
Next,
\begin{eqnarray}
& & \mu \langle f(\mathbf{x}, t) \mathcal{L}X(\mathbf{y}, t) \rangle = \nonumber \\ & &  \:\:\:\:\:\:\:\:\:\:\:\: \mathcal{L}_y \langle f(\mathbf{x}, t) X(\mathbf{y}, t) \rangle = \mathcal{L}_yD(\mathbf{x}, \mathbf{y}),  \nonumber
\end{eqnarray}
or, substituting for $D(\mathbf{x}, \mathbf{y})$ from the fluctuation dissipation relation (42),
\begin{equation}
\mu \langle f(\mathbf{x}, t) \mathcal{L}X(\mathbf{y}, t) \rangle = \frac{\mu}{2} \mathcal{L}_y(\mathcal{L}_x + \mathcal{L}_y)M(\mathbf{x}, \mathbf{y}). 
\end{equation}
Similarly,
\begin{equation}
\mu \langle f(\mathbf{y}, t) \mathcal{L} X(\mathbf{x}, t) \rangle = \frac{\mu}{2}\mathcal{L}_x(\mathcal{L}_x + \mathcal{L}_y)M(\mathbf{x}, \mathbf{y}).  
\end{equation}
From (A3) and (A4),
\begin{equation}
\mu \langle f(\mathbf{x}, t) \mathcal{L}X(\mathbf{x}, t) \rangle = \frac{\mu}{4} (\mathcal{L}_x + \mathcal{L}_y)^2 M(\mathbf{x}, \mathbf{x}).  
\end{equation}
Substituting (A2) and (A5) into (A1),
\begin{equation}
\langle \Dot{q} - \Dot{w} \rangle = \mu \left(\mathcal{L}_xL_y - \frac{1}{4}(\mathcal{L}_x + \mathcal{L}_y)^2 \right)M(\mathbf{x}, \mathbf{x}).   \nonumber
\end{equation}
Here,
\begin{equation}
\mathcal{L}_x \mathcal{L}_y - \frac{1}{4}(\mathcal{L}_x + \mathcal{L}_y)^2 = -\frac{1}{4}(\mathcal{L}_x - \mathcal{L}_y)^2.  \nonumber
\end{equation}
so
\begin{equation}
\langle \Dot{q} - \Dot{w} \rangle = -\frac{\mu}{4} (\mathcal{L}_x - \mathcal{L}_y)^2M(\mathbf{x}, \mathbf{x}),  \nonumber
\end{equation}
or finally,
\begin{equation}
\langle \Dot{q} - \Dot{w} \rangle =  \frac{1}{2} (\mathcal{L}_x - \mathcal{L}_y) \Omega(\mathbf{x}, \mathbf{x}). \nonumber
\end{equation}
\medskip 


\bibliography{Active}

\end{document}